# How inter-rater variability relates to aleatoric and epistemic uncertainty: a case study with deep learning-based paraspinal muscle segmentation

Parinaz Roshanzamir[1], Hassan Rivaz[1], Joshua Ahn[2], Hamza Mirza[2], Neda Naghdi[3], Meagan Anstruther[3], Michele C. Battié[4], Maryse Fortin[3], Yiming Xiao[5]

[1] Department of Electrical and Computer Engineering, Concordia University, Montreal, Canada
`parinaz.roshanzamir@concordia.ca`
[2] Faculty of Health Sciences, Western University, London, Canada
[3] Health, Kinesiology, and Applied Physiology, Concordia University, Montreal, Canada
[4] School of Physical Therapy and Western's Bone and Joint Institute, Western University, London, Canada
[5] Department of Computer Science and Software Engineering, Concordia University, Montreal, Canada

**Abstract.** Recent developments in deep learning (DL) techniques have led to great performance improvement in medical image segmentation tasks, especially with the latest Transformer model and its variants. While labels from fusing multi-rater manual segmentations are often employed as ideal ground truths in DL model training, inter-rater variability due to factors such as training bias, image noise, and extreme anatomical variability can still affect the performance and uncertainty of the resulting algorithms. Knowledge regarding how inter-rater variability affects the reliability of the resulting DL algorithms, a key element in clinical deployment, can help inform better training data construction and DL models, but has not been explored extensively. In this paper, we measure aleatoric and epistemic uncertainties using test-time augmentation (TTA), test-time dropout (TTD), and deep ensemble to explore their relationship with inter-rater variability. Furthermore, we compare UNet and TransUNet to study the impacts of Transformers on model uncertainty with two label fusion strategies. We conduct a case study using multi-class paraspinal muscle segmentation from T2w MRIs. Our study reveals the interplay between inter-rater variability and uncertainties, affected by choices of label fusion strategies and DL models.

**Keywords:** Inter-rater variability, Uncertainty, TransUNet, Segmentation

## 1 Introduction

In recent years, deep learning (DL) techniques have shown remarkable success in various fields, including medical domains. However, using DL models in safety-critical applications, such as medical diagnosis and treatment, requires not only high accuracy but also a proper understanding of the model's uncertainty, which is crucial for the safety and adaptability of medical DL algorithms. In general, a DL model's uncertainty can be classified into two main categories: epistemic and aleatoric [4]. Epistemic uncertainty is related to the model's lack of knowledge about the data and can be reduced by collecting more data or optimizing the model's architecture and training process. On the other hand, aleatoric uncertainty is related to inherent factors of the data (e.g., noise) and cannot be reduced. Additionally, in image segmentation tasks, where supervised learning is widely used, another important factor that can affect uncertainty is inter-rater variability. Supervised segmentation algorithms require training data with well-annotated ground truth (GT) masks. While GT masks obtained by fusing annotations of multiple experts are commonly recommended, it is still costly and the best practice to combine different annotations is still being explored. Various factors can affect inter-rater variability in manual medical image segmentation, including differences in expertise, rater style [18], image noise, extreme anatomical variations among individuals, and so on. In turn, inter-rater variability propagates the influences of these factors to the resulting DL models through training as uncertainties of the algorithms. For example, image noise and measurement errors (e.g., due to partial volume effects) can result in aleatoric uncertainty while extreme individual anatomical variations, which may not be sufficiently represented in the data, can contribute to epistemic uncertainty. Therefore, knowledge regarding the relationship of inter-rater variability with aleatoric and epistemic uncertainties in DL models can help better understand their performance and reliability and inform the dataset design and learning strategies to improve them.

To date, various methods have been proposed to measure aleatoric and epistemic uncertainties of DL models. In terms of epistemic uncertainty, Bayesian DL estimates a distribution for each weight in the network and uses these distributions to measure epistemic uncertainty [10,20]. As Bayesian networks can bear high computational costs, more efficient approaches have been reported for uncertainty estimation. Gal and Ghahramani [6] proposed dropout at test-time to approximate Bayesian neural networks for uncertainty estimation while deep ensemble trained multiple versions of the same DL model to derive epistemic uncertainty [11]. Finally, variational inference [9] has also been used but limits the types of DL models in application. Different approaches have been reported ever since to improve the quality of the uncertainties obtained from these methods [12]. Another important aspect in accurate uncertainty quantification is the metric. Camarasa et al. [1] performed an extensive study on different metrics for measuring epistemic uncertainty using test-time dropout (TTD). They concluded that the measure of entropy produces uncertainty maps that are in correspondence with the misclassification in the model. Utilizing a sampling strategy similar to TTD, Wang et al. [19] proposed test-time augmentation (TTA) for aleatoric uncertainty assessment, which samples from the data distribution by using input data augmentation at inference time.

In training data, inter-rater variability can be measured as the entropy of the rater annotations. Lemay et al. [13] showed the superiority of random sampling and STAPLE in inter-rater variability preservation and image segmentation accuracy with a UNet. Jensen et al. [8] discovered that random sampling leads to higher classification accuracy and better-calibrated results. Vincent et al. [18] characterized rater style in terms of bias and variance of raters' annotations and explored the relationship be-tween rater bias and data uncertainty. Nichyporuk et al. [15] proposed a segmentation model that can learn the bias in the annotations for better results. However, previous studies haven't explored the impact of label fusion methods on aleatoric and epistemic uncertainties, or the potential relationship between inter-rater variability and these uncertainties. In addition, most of them only used a conventional UNet as the base model. In recent years, Transformers that better model long-range dependencies via self-attention have gained popularity in vision tasks [5], and the hybrid Transformer-convolutional neural network (CNN) models that complement their merits, such as TransUNet [2], have shown better segmentation accuracy [16]. With a different mechanism, the addition of Transformers can have potential effects on the uncertainty of a segmentation model, which has not been investigated, but can be highly valuable.

To address the aforementioned knowledge gaps, In this study, we investigate inter-rater variability in relation to DL model uncertainties using commonly employed TTA, TTD, and deep ensemble techniques for MRI-based paraspinal muscle segmentation, with a comparison of UNet and TransUNet. Our work has three main novel contributions: First, we are the first to compare Transformers and CNNs for their impacts on model uncertainties and the encoding of inter-rater variability, especially in a multi-class segmentation setting. Second, we explore the effect of label fusion methods during network training on DL model uncertainty (aleatoric and epistemic) for the first time. Lastly and most importantly, we explore the relationship between inter-rater variability and aleatoric/epistemic uncertainties, which has not been done so far. We hope the article will offer instrumental insights to facilitate the design and selection of DL datasets, training strategies, and model architectures.

## 2      Materials and Methodology

### 2.1    Inter-rater Variability

To measure inter-rater variability, we use the entropy of the average GT annotations for each image [13]. The pixel-wise entropy is calculated as:

$$H(y_i) = -\sum_{c=1}^{C} P(y_i = c) \log(P(y_i)) \tag{1}$$

where C is the number of classes, and $y_i$ is the GT annotation at voxel *i*, obtained by simply averaging the one-hot GT masks from all raters. Similarly, the entropy of the model predictions can be calculated as the entropy of the softmax layer outputs before binarization. This entropy can be considered as the prediction inter-rater variability. *An ideal model should preserve the inter-rater variability during inference and produce prediction entropies similar to the GT entropy.* To quantify the perseverance of inter-rater variability in a model, we calculated the class-wise Brier score [13] as:

$$Brier\ Score\ = \frac{1}{N_{image}} \sum_{k=1}^{N_{image}} \left( \frac{1}{N_{voxel}} \sum_{i=1}^{N_{voxel}} (y_{i,k} - \widehat{y_{i,k}})^2 \right) \tag{2}$$

where $N_{image}$ is the total number of images in the test set, $N_{voxel}$ is the number of voxels in each image, $y_{i,k}$ is the average GT, and $\widehat{y_{i,k}}$ is the prediction for voxel *i* of image *k* (the softmax probability). A Brier score close to zero indicates perfect preservation of inter-rater variability in model predictions [13].

### 2.2 Aleatoric and Epistemic Uncertainty Assessment

In this study, we compare TTD and deep ensemble for epistemic uncertainty assessment, and use TTA for measuring aleatoric uncertainty for both the UNet and TransUNet. In each experiment, we acquire 10 samples. In TTA and TTD, these samples are obtained through 10 forward passes through the model for each image, and in deep ensembles, each image is fed to 10 independently trained models of the same architecture. The obtained samples are then averaged to produce a final prediction for each image. This prediction is then used to calculate the entropy based on Eq. 1. To compare the quality of the resulting uncertainty maps of TTD and deep ensembles, we measure the association between uncertainties and misclassifications, using the framework provided by Mobiny et al. [14]. Conventionally, in a confusion matrix the term "positive" is used for "capturing a target label/class", but here, we define it as "capture a high uncertainty". An ideal model should be uncertain only when making a wrong prediction. Thus, a "false positive" means that the model is highly uncertain about a correct classification. Identifying a voxel as "uncertain" requires thresholding an uncertainty map. We perform the operation at multiple values, calculate the corresponding precision and recall metrics, and finally use the area under the precision-recall curve (AUC-PR) as an indicator of the quality of an uncertainty map.

### 2.3 Network Architectures and Label Fusion

Lemay et al [13] showed that the method used for label fusion in a multi-rater dataset can affect the model calibration and how well it preserves inter-rater variability. In this study, we used two different methods for integrating multi-rater annotations into our training framework: 1) The majority vote of all raters is used as the GT for each image; 2) The annotation of one rater is randomly selected at each epoch during training (refer to as random sampling). In order to explore the impact of self-attention modules on uncertainty, we train two sets of models, using TransUNet and UNet as the base model architectures. The TransUNet model has four upsampling layers with two convolution blocks at each layer. For a fair comparison, we used a UNet model with the same number of layers as the TransUNet (4 layers). With these two architectures and two label fusion methods, we trained four models for TTA and TTD-related analysis. To achieve deep ensembles for measuring epistemic uncertainty, we also trained a set of 10 UNets and a set of 10 TransUNets with majority vote GT.

### 2.4 Dataset

Our dataset consists of a total of 673 lumbosacral T2-weighted (T2w) MR images of 119 patients (59 male, age=30~59y) from the European research consortium project, Genodisc, on commonly diagnosed lumbar pathologies. The subjects were selected with the factors of sex and age roughly equally distributed. Our study was approved by local research ethics board. The MRI scans are from 6 different disc levels. However, due to imaging artifacts and cropping, not all patients have usable axial slices at all spinal levels. All axial MR images were processed with non-local means denoising [3] and N4 inhomogeneity correction [17] to improve image quality. Then, the left and right multifidus (MF) and erector spinae (ES) muscles were manually segmented for all patients independently by three different raters to study low back pain [21], resulting in four segmentation classes for each image (see Fig. 1). All raters had two training sessions to ensure the quality and consistent protocol of the segmentation.

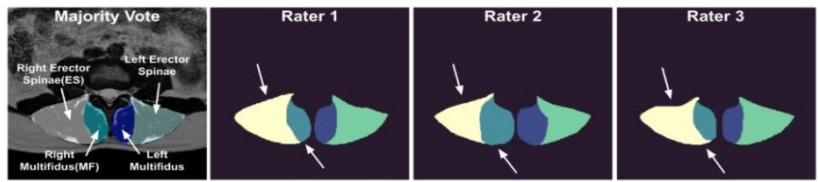

**Fig. 1.** Left to right: axial MRI of paraspinal muscles, along with the majority vote label and the individual rater annotations. The arrows indicate the differences among rater annotations.

## 3 Experiments and Results

### 3.1 Experimental Set-up and Implementation Details

Each model is trained to segment four paraspinal muscles. We divide the dataset subject-wise into training, validation, and test sets with respect to age group and sex. Specifically, 24 subjects are selected for testing, 76 for training, and 19 for validation. For TTD, we add a dropout layer to each of the convolution blocks in the upsampling layers of UNet and TransUNet, resulting in the addition of a total of 8 dropouts in each model. Both models are trained for 250 epochs with early stopping applied to avoid overfitting. We applied random rotation, translation, intensity shift, and Gaussian noise during the training and applied the same augmentations at test time for TTA [19]. As mentioned in the previous section, we use entropy for measuring uncertainties in this study (Eq. 1). In order to explore the correlation between aleatoric and epistemic uncertainties with inter-rater variability, we calculate the Pearson correlation coefficients and conduct a variance partitioning analysis to verify the percentage of variance explained by the uncertainties for inter-rater variability (i.e., GT entropy).

### 3.2 Results

The Brier scores for inter-rater variability preservation assessment are listed in Table 1 for all the models and all segmented muscle groups. Additionally, we also plotted the prediction entropy against GT entropy with the associated correlations in Fig. 2(a). These results indicate that compared with UNet, TransUNet better preserves inter-rater variability with lower Brier scores and higher "prediction entropy vs. GT entropy" correlations. Also, we observe that training with random sampling results in lower Brier scores than using majority vote for the same models. To evaluate the quality of epistemic uncertainty estimation, the AUC-PR results are detailed in Table 2, where TransUNet outperforms its UNet counterpart in both TTD and deep ensemble approaches. Here, random sampling, TTD, and deep ensemble show similar performance so for the rest of the experiments to compare random sampling and majority vote, we only trained the deep ensemble model with majority vote GT. Table 3 contains the correlations of inter-rater variability with aleatoric and epistemic uncertainties, and the evidence shows that in our case study, inter-rater variability is more strongly associated with epistemic uncertainty than the aleatoric one, and the phenomenon is stronger for TransUNet. The superiority of TransUNet is further demonstrated in Table 4, where the models' performance is evaluated with Dice score. According to the scatter plots of Fig. 2(b), we also see that higher correlation is produced with random sampling training, and the UNet model contains higher epistemic uncertainties. Finally, from the variance partitioning analysis (see Table 5), we observe that epistemic uncertainty accounts for ~35% of the GT entropy variance with TransUNet and ~12% with UNet. Additionally, we observe that aleatoric uncertainty only explains a very small portion of the variance.

**Table 1.** Quantitative assessment of inter-rater variability preservation in the trained models.

|  | Average Brier Score ($\times 10^{-3}$) | | | |
| --- | --- | --- | --- | --- |
|  | Right MF | Left MF | Right ES | Left ES |
| TransUNet (majority vote) | 1.324 | 1.209 | 1.936 | 1.890 |
| UNet (majority vote) | 2.071 | 1.973 | 3.286 | 3.146 |
| TransUNet (random sampling) | 1.179 | 1.103 | 1.742 | 1.758 |
| UNet (random sampling) | 1.769 | 1.670 | 2.866 | 2.912 |

**Table 2.** AUC-PR for epistemic uncertainty. Each column shows a method for measuring the uncertainty and the training method, while the rows indicate the utilized models.

|  | AUC-PR | | |
| --- | --- | --- | --- |
|  | TTD-majority vote | TTD-random sampling | Deep Ensemble |
| TransUNet | 0.3753 | 0.3737 | 0.3831 |
| UNet | 0.3387 | 0.3337 | 0.3265 |

**Table 3.** Correlation of epistemic and aleatoric uncertainties with inter-rater variability. Majority vote is shown as "Maj" while random sampling is shown as "Rand".

|  | Pearson Correlation Coefficient | | | | |
| --- | --- | --- | --- | --- | --- |
|  | TTD – Maj | TTD - Rand | Deep Ensemble | TTA - Maj | TTA -Rand |
| TransUNet | 0.5874 | 0.6165 | 0.5985 | 0.0500 | 0.1028 |
| UNet | 0.3380 | 0.3635 | 0.3574 | 0.0719 | 0.0163 |

## 4   Discussion

With the experiments, the results of the Brier scores and "prediction entropy vs. GT entropy" correlations indicate that both the TransUNet architecture and random sam-pling have positive impacts on preserving inter-rater variability. Furthermore, as Fig. 2(b) shows, TransUNet produces lower epistemic uncertainty with tighter distribution. Our results confirm the conclusion of Lemay et al. [13] on the benefit of random sampling in preserving inter-rater variability. Furthermore, we observe that it also results in higher prediction entropies (Fig. 2(a)) with more uncertain results. When assessing AUC-PR, TransUNet offers a better quality of epistemic uncertainty estimation while the advantages of different estimation techniques and training strategies are not clear. When comparing the correlations of inter-rater variability (i.e., GT entropy) with aleatoric and epistemic uncertainties, the results in Table 3 demonstrate that epistemic uncertainty has a stronger association in our segmentation task and database while no significant correlations with aleatoric uncertainties were found, regardless of the DL model choice. This may be partially explained by the fact that manual segmentations were performed based on pre-processed images with similar noise levels, reducing the chance of inter-rater variability being affected by image noise. Future studies to explore the impact of image noise levels and artifacts (e.g., bias fields) can further verify this hypothesis, but require a more extensive and costly experimental set-up with human raters. In addition, there is also a higher correlation between inter-rater variability and epistemic uncertainty with TransUNet and random sampling as shown in Table 3, proving that model uncertainty can be network-dependent and better preservation of inter-rater variability leads to a stronger link to the model uncertainty. Although better "epistemic uncertainty vs. inter-rater variability" correlation and preservation of inter-rater variability are desirable as they can result in more effective reduction of uncertainty through lowering inter-rater variability, the overall higher prediction entropy and epistemic uncertainty may be a price to pay in the case of random sampling compared to majority vote. As a final evaluation, we used variance partitioning (Table 5) to quantify the contributions of aleatoric and epistemic uncertainties toward inter-rater variability. This way, we leverage the DL models to understand the source of inter-rater variability, which is difficult to quantify from human raters [7]. The results indicate a partial influence of epistemic uncertainty that may be due to the factors of anatomical variability (common in pathological paraspinal muscles) and difference in visual perception, and minimum contribution from aleatoric uncertainty. This suggests the benefit of preprocessing and systematic training for expert labeling. For all experiments, the incorporation of Transformers has positive impacts in lowering the uncertainty and encoding inter-rater variability, potentially leading to better segmentation accuracy. Their ability to encode long-range content over the image may play a key role in the observed behaviors.

## 5    Conclusion

In this paper, we explored the relationship of inter-rater variability with aleatoric and epistemic uncertainties, using two DL models and two label fusion methods. Our case study indicated that inter-rater variability has a high correlation with epistemic uncertainty and no significant correlation with aleatoric uncertainty. Moreover, we showed that TransUNet better preserves inter-rater variability and its correlation with epistem-ic uncertainty, and it also has lower epistemic uncertainty and prediction entropy than UNet, potentially explaining its segmentation accuracy. Finally, our results showed that the label fusion method not only affects the preservation of inter-rater variability but it also affects epistemic uncertainty as well.

**Table 4.** Model performance measured by Dice Score. The superior performance of TransUNet compared to the UNet counterpart is indicated by **(p<0.01) and *(p<0.05).

|  | Average Dice Score (%) | | | |
| --- | --- | --- | --- | --- |
|  | Right MF | Left MF | Right ES | Left ES |
| TransUNet (majority vote-TTD) | **94.36±2.74 | **94.77±2.37 | **94.29±3.35 | **94.18±3.77 |
| UNet (majority vote-TTD) | 92.41±8.9 | 92.77±8.52 | 92.21±9.52 | 92.06±9.31 |
| TransUNet (random sampling-TTD) | 94.15±3.14 | 94.44±2.53 | *94.08±3.38 | *93.87±3.94 |
| UNet (random sampling-TTD) | 92.76±8.86 | 93.07±8.67 | 92.49±9.06 | 91.92±10.04 |
| TransUNet (ensemble) | *94.72±2.38 | *94.80±2.49 | **94.75±3.26 | **94.50±3.18 |
| UNet (ensemble) | 92.83±8.73 | 93.23±8.56 | 92.79±8.96 | 92.05±10.12 |

**Table 5.** Variance partitioning analysis for inter-rater variability. The values show the percentage of inter-rater variability variation related to the epistemic and aleatoric uncertainties.

|  |  | Epistemic (%) | Aleatoric (%) | Joint (%) |
| --- | --- | --- | --- | --- |
| TransUNet | TTA+TTD(majority vote) | 34.506 | 0.251 | 34.896 |
|  | TTA+TTD(random sampling) | 38.007 | 0.251 | 38.360 |
|  | TTA+Ensemble | 35.803 | 0.251 | 35.986 |
| UNet | TTA+TTD(majority vote) | 11.422 | 0.517 | 11.517 |
|  | TTA+TTD(random sampling) | 12.772 | 0.517 | 12.919 |
|  | TTA+Ensemble | 13.215 | 0.517 | 13.385 |

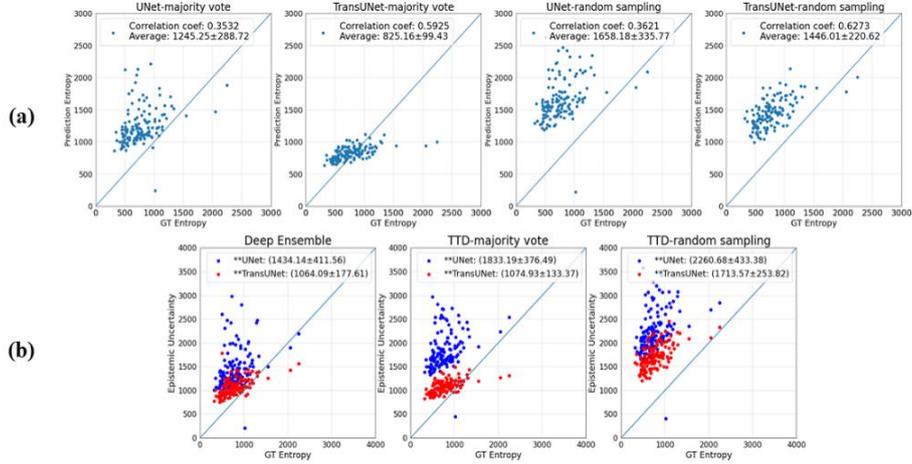

**Fig. 2.** (a) Assessment of preservation of inter-rater variability, along with the average entropy and Pearson correlation coefficient shown in the graphs. (b) Comparison of epistemic uncertainty with inter-rater variability, along the average uncertainties shown in the graphs. Significant correlation is denoted by **(p<0.01).

## 6      Acknowledgment

We acknowledge the support of the Natural Sciences and Engineering Research Council of Canada (NSERC) and NVIDIA for donation of the GPU.